\documentclass[final,twocolumn]{elsarticle}

\usepackage[a4paper,margin=2cm]{geometry}
\usepackage{float}
\usepackage{graphicx}
\usepackage{amssymb}
\usepackage{multirow}
\usepackage{lineno}
\usepackage{booktabs}
\usepackage{threeparttable}
\usepackage{caption}
\usepackage[version=3]{mhchem} % Formula subscripts using \ce{}
\usepackage{amsmath}

\journal{Journal of Molecular Biology}

\begin{document}

\begin{frontmatter}

\title{Confinement effects on protein stability in a freezing water environment}

\author[a,b]{Yanis R. Espinosa}
\author[b,c,d]{H. Ariel Alvarez}
\author[b,e]{C. Manuel Carlevaro}

\address[a]{Faculty of Basic Sciences, Research Group in Molecular Biology and Genetics (BIOMOGEN), University of Pamplona, Pamplona, Colombia.}
\address[b]{Instituto de Física de Líquidos y Sistemas Biológicos (CONICET-UNLP), Calle 59 Nro 789, 1900 La Plata, Argentina.}
\address[c]{Instituto de Ciencias de la Salud, Universidad Nacional Arturo Jauretche, Florencio Varela, Argentina.}
\address[d]{Departamento de Ciencias Biológicas, Facultad de Ciencias Exactas, Universidad Nacional de La Plata, La Plata, Argentina.}
\address[e]{Universidad Tecnológica Nacional, Facultad Regional La Plata, Centro de Investigación en Mecánica Experimental y Computacional, Av. 60 esq. 124 s/n, 1923 Berisso, Argentina.}

\begin{abstract} Understanding how proteins behave at low temperatures remains a central challenge in biophysics, with direct implications for cold denaturation and cryopreservation. While cold denaturation of proteins in the supercooled liquid regime has been studied extensively, the behavior of a protein embedded in a growing ice lattice remains largely inaccessible to experiments. Here we use molecular dynamics simulations that explicitly capture ice $I_h$ formation to characterize the conformational dynamics of yeast frataxin (Yfh1) as its aqueous environment crystallizes. Using four independent ice-seeded replicas and liquid-water controls at three temperatures, we first validate the liquid–solid transition through convergent changes in solvent density, potential energy, and local bond-order parameters (W4, W6). Principal component analysis (PCA), dihedral PCA (dPCA), and free-energy landscapes then reveal that crystallization of the solvent markedly reshapes the accessible conformational space, shifting it from a continuous, highly connected regime in liquid water toward a discretized landscape dominated by confined states. Complementary analyses of solvent-accessible surface area (SASA), radius of gyration, and hydrogen bonding indicate a solvent-driven reorganization of protein–water interactions: although first-shell water remains liquid-like, its surface density increases under freezing, while conformational sampling contracts. Together, these results indicate that protein behavior at low temperatures is governed not by temperature alone but by the structural organization of the surrounding water. By imposing geometrical constraints on the solvent, ice formation restricts conformational sampling while preserving—and even densifying—the interfacial hydration layer, highlighting the role of water structure as a determinant of protein stability under freezing conditions.
\end{abstract}

\end{frontmatter}

%%%END OF TITLE, AUTHORS AND ABSTRACT%%%

%%%MAIN TEXT%%%%

%%%FONT SETUP - please do not change any commands within this section
\renewcommand*\rmdefault{bch}\normalfont\upshape
\rmfamily

\section{Introduction}

Since the mid-20\textsuperscript{th} century, there has been considerable interest in understanding the effect of low temperatures on protein stability. One of the earliest reports on this phenomenon was presented by Simpson and Kauzmann \cite{simpson1953}, who showed that ovalbumin is more resistant to urea-induced denaturation at low temperatures. Decades later, Privalov provided a comprehensive thermodynamic explanation for cold denaturation, proposing that it is driven by the favorable hydration of non-polar groups on the protein surface at low temperatures \cite{privalov1990}. According to his model, the enthalpy of hydration becomes increasingly negative as temperature decreases, making the exposure of hydrophobic groups to water more favorable than their burial in the protein core. As a result, the hydrophobic effect—normally a major contributor to protein stability—diminishes, since the entropic cost of organizing water around hydrophobic groups is offset by a larger enthalpic gain.

This phenomenon presents a thermodynamic paradox. The hydrophobic effect is traditionally understood as the tendency of non-polar residues to cluster in the protein core due to the entropic penalty associated with structuring water molecules around them. Given that water forms more ordered hydrogen-bond networks at lower temperatures, one might expect the hydrophobic effect to strengthen, thereby enhancing protein stability \cite{tsai2002}. However, experimental evidence shows that many proteins undergo cold denaturation at low temperatures, indicating that the unfolded state becomes thermodynamically favored over the native structure \cite{sosnick2025}.

In most mesophilic proteins, cold denaturation occurs at temperatures well below the freezing point of water (0~$^{\circ}$C), making it difficult to study them under physiological conditions \cite{temussi2025}. Remarkably, the yeast frataxin protein (Yfh1 of \emph{Saccharomyces cerevisiae}) exhibits cold denaturation at temperatures near 5~$^{\circ}$C and at neutral pH, without the need for chemical denaturants, extreme pressures, or destabilizing mutations. This unique behavior allows for the direct determination of stability curves ($\Delta G$ vs. $T$) under near-physiological conditions \cite{adrover2010, adrover2012, sanfelice2016}.

Studies on Yfh1 have provided the first direct experimental support for Privalov's theory, demonstrating a significant increase in hydration of the cold-denatured state, characterized by a greater number of protein–water hydrogen bonds \cite{adrover2012, puglisi2021, temussi2025}. Furthermore, cold and heat denaturation mechanisms are not symmetric: the cold-denatured state of Yfh1 is more expanded and more hydrated than its heat-denatured counterpart \cite{sanfelice2015, puglisi2022, alfano2017}.

Water is widely recognized as a key determinant of protein stability \cite{campanile2024, hishida2022}. Although structural studies of Yfh1 have shown that cold-induced unfolding is thermodynamically driven by favorable hydration \cite{adrover2012, puglisi2021}, little is known about how the hydration network reorganizes as freezing begins, and how it drives or not protein denaturation.

Muller proposed that hydrogen bonds (HBs) in the hydration shell (hs) of solutes are enthalpically stronger than those in bulk (b) water ($\Delta H_{\mathrm{hs}} > \Delta H_{\mathrm{b}}$), but also entropically more fragile ($\Delta S_{\mathrm{hs}} > \Delta S_{\mathrm{b}}$), making them more susceptible to breaking \cite{muller1990, graziano2005}. Graziano further emphasized that the primary driving force of cold denaturation is the excluded volume effect of the solvent \cite{graziano2014}, as the reorganization of water–water HBs leads to minimal changes in Gibbs free energy, characterized by enthalpy–entropy compensation \cite{riccio2011, graziano2014}.

Dias and colleagues \cite{dias2010, dias2012} proposed that at low temperatures, the fraction of broken HBs in the hydration shell ($f_{\mathrm{hs}}$) may be lower than in bulk water ($f_{\mathrm{b}}$), favoring the formation of clathrate-like structures or solvent-separated configurations (SSC) around the protein. These ordered water cages would release heat upon formation, potentially contributing to cold denaturation. 

Crucially, all of these mechanisms—favorable hydration of non-polar groups, the enthalpic and entropic asymmetry of hydration-shell hydrogen bonds, excluded-volume effects, and clathrate-like ordering—have been formulated for the supercooled \textit{liquid} regime, where water remains disordered. How the hydration network reorganizes once the solvent actually crystallizes, and whether this crystallization drives, suppresses, or merely constrains protein conformational change, remains largely unexplored.

In this work, we present a molecular-level analysis of the hydration and conformational behavior of yeast frataxin (Yfh1) as its aqueous environment undergoes a liquid–solid transition. Using molecular dynamics simulations that explicitly capture ice $I_h$ growth, we compare freezing conditions against liquid-water controls at room temperature (\emph{Ta}), at the temperature of maximum density (\emph{TMD}), and near the melting point (\emph{Tm}), thereby disentangling the effect of temperature itself from that of solvent phase and the anomalous properties of water below \emph{TMD}. Our central question is whether the reorganization of the water network upon crystallization modulates protein–water interactions and conformational dynamics beyond what temperature alone would predict. We hypothesize that ice formation acts primarily as a geometrical constraint on the solvent: rather than driving global unfolding, it should restrict the accessible conformational space while reshaping—not simply depleting—the interfacial hydration layer. By accessing a regime in which the protein is embedded within a growing ice lattice, a scenario that remains largely inaccessible experimentally, we aim to clarify how solvent structure governs protein stability under freezing conditions.

\section{Computational Methods}

To investigate how the hydration shell surrounding a protein responds to decreasing temperature, we implemented the model proposed by Kuiper et al. \cite{kuiper2015}, which simulates continuous ice growth on an infinite plane using periodic boundary conditions (PBC). This model employs a crystallization seed composed of 1,692 water molecules with fixed Cartesian coordinates, arranged in a prismatic hexagonal ice lattice ($I_h$). This configuration promotes freezing of the surrounding water when the system temperature drops below the melting point (Tm) of the TIP4P water model \cite{jorgensen1983}.
We constructed a simulation box with dimensions X = 5.20 nm, Y = 12.67 nm, and Z = 9.10 nm, containing the following components: (i) a prismatic ice seed ($I_h$); (ii) 17,130 randomly distributed water molecules; (iii) the apo form of \emph{Saccharomyces cerevisiae} frataxin (PDB code: 2GA5) \cite{he2004}; and (iv) fifteen Na$^+$ counterions for charge neutralization. We selected the TIP4P/2005 water model \cite{abascal2005} due to its accurate prediction of water density at 1 bar (maximum density $\rho = 1.0005$ g/cm$^{3}$ at 278.0 K). Although its melting point is~23 °C below the experimental value (249.5 K ± 0.5 K), it reliably reproduces key properties such as the enthalpy of fusion, hydrogen bonding, and the structural characteristics of ice \cite{conde2017, blazquez2022}.

Four replicas of the system (\emph{R1}, \emph{R2}, \emph{R3}, and \emph{R4}) were generated to analyze ice growth around the protein. Three control systems without an ice seed were constructed: \emph{Ta}, \emph{TMD}, and \emph{Tm}. Specifically, \emph{Ta} was simulated at room temperature (298.0 K), while \emph{TMD} was simulated at the maximum density temperature (278.0 K) of the TIP4P/2005 model. These systems remained in a liquid state throughout the simulations.  \emph{Tm} system and the four replicas \emph{R1-R4} were initially simulated at the melting temperature (249.5 K) of the TIP4P/2005 model for 0.5 $\mu$s, and then continued for another 0.5 $\mu$s at a slightly lower temperature (247 K), as summarized in Table \ref{Table:1}. 

\begin{table}[ht!]
\centering
\begin{threeparttable}
\caption{\textbf{Simulation systems.}}
\label{Table:1}
\begin{tabular}{lccccc}
\toprule
\textbf{Systems} & \textbf{Seed} & \textbf{Waters} & \textbf{$T_i$ (K)} & \textbf{$T_f$ (K)} & \textbf{Phase} \\
\midrule
Ta        & No  & 19\,240  & 298.0 & 298.0 & Liquid \\
TMD       & No  & 19\,240  & 278.0 & 278.0 & Liquid \\
Tm        & No  & 19\,240  & 249.5 & 247.0 & Liquid \\
R1--R4    & Yes & 18\,822  & 249.5 & 247.0 & Ice $I_h$ \\
\bottomrule
\end{tabular}
\begin{tablenotes}
\footnotesize
\item[$T_i$] Initial temperature
\item[$T_f$] Final temperature
\end{tablenotes}
\end{threeparttable}
\end{table}

Additionally, another control system was simulated for 0.5 $\mu$s, which was composed only of 19,427 TIP4P/2005 water molecules with an ice $I_h$ seed (\emph{Tm}$_{seed}$) at 247.0 K.

\subsection{Stabilization Process}
All systems (\emph{Ta}, \emph{TMD}, \emph{Tm}, \emph{Tm}$_{seed}$, and \emph{R1–R4}) underwent energy minimization using the Steepest Descent method for 5,000 steps or until the maximum force dropped below 10 kJ mol$^{-1}$ nm$^{-1}$. During this stage, protein backbone atoms were positionally restrained with a harmonic potential (force constant: 1000 kJ mol$^{-1}$ nm$^{-2}$), gradually reduced to zero following the equilibration steps detailed in Supplementary Table 1S.
Temperature and pressure were controlled in the $NpT$ ensemble using the velocity rescaling thermostat \cite{bussi2007} (coupling constant $\tau = 0.1$ ps) and the Berendsen barostat \cite{berendsen1984} (coupling constant $\tau = 0.5$ ps). Given the orientation of the $I_h$ seed along the Z-axis, we applied semi-isotropic pressure coupling, as described by Pereyra et al. \cite{pereyra2011}. Compressibility was set to $4.5 \times 10^{-5}$ bar$^{-1}$ in the XY plane and $4.5 \times 10^{-7}$ bar$^{-1}$ along the Z-axis. Molecular dynamics (MD) equilibration trajectories were generated for 1 ns with a time step of 1 fs. 

\subsection{Production Simulations}
Production MD simulations were conducted for 1 $\mu$s with a 2 fs integration time step. The Nose–Hoover thermostat \cite{evans1985} ($\tau = 1.0$ ps) and Parrinello–Rahman barostat \cite{parrinello1981} ($\tau = 5.0$ ps) were used to maintain temperature and pressure, preserving the same semi-isotropic pressure coupling as in the equilibration phase.

Positional restrictions on the protein were removed during production, except for the oxygen atoms in the seed $I_h$, which remained constrained (force constant: 5000 kJ mol$^{-1}$ nm$^{-2}$) in replicas \emph{R1–R4} and \emph{Tm}$_{seed}$. 

Electrostatic interactions were treated with the particle-mesh Ewald (PME) method \cite{essmann1995, harvey2009}, using a 0.12 nm grid spacing. Cutoff distances for Coulomb and van der Waals interactions were set to 1.0 nm. The lengths of the bonds involving the protein atoms were constrained using the LINCS algorithm \cite{hess1997}.

All simulations used the Amber03w force field \cite{best2010} and were performed with \texttt{GROMACS} 2022.6 \cite{abraham2015}. The Computations were carried out on a Linux workstation equipped with an Intel Core i7–6700 (3.40 GHz, 8 cores) and an \texttt{NVIDIA} GeForce GTX 1080 GPU.

In addition to the usual analyses of this type of systems (SASA, Hydrogen bonds, Radius of gyration, and energetic variables), we conducted a study on how the freezing environment remodels the conformational landscape of the protein.

\subsection{Dihedral Principal Component Analysis (dPCA)}

To characterize the free energy landscape and conformational transitions of yeast frataxin (Yfh1) under supercooled conditions and ice formation, a dihedral-based principal component analysis (dPCA) was performed \cite{mu2005energy, altis2007}. Since the N-terminal region of Yfh1 (residues 1--19) is intrinsically disordered and exhibits stochastic fluctuations that artificially increase the apparent conformational variability, this region was excluded from all structural analyses. This allowed the study to focus exclusively on the conformational dynamics of the globular domain (residues 20 onward).

The analysis was carried out following the workflow implemented in the \texttt{GROMACS} 2022.6 package, using the 1~\textmu s production trajectories corresponding to the liquid-phase systems (\emph{Ta}, \emph{TMD}, and \emph{Tm}) and to the ice-seeded replicas (\emph{R1--R4}). Backbone dihedral angles ($\phi$ and $\psi$) were extracted using the \texttt{gmx angle} tool. To properly address the inherent statistical circularity and periodicity of angular variables, each dihedral angle $\alpha_n$ was mapped onto a linear metric space using its sine and cosine components, according to:
\[
q_{2n-1} = \cos \alpha_n, \qquad q_{2n} = \sin \alpha_n
\]
following previously established procedures \cite{mu2005energy, altis2007}. The covariance matrix was then constructed from the transformed data using \texttt{gmx covar} and subsequently diagonalized to obtain the eigenvalues and eigenvectors, ranked in descending order according to the variance captured. Finally, the trajectories were projected onto the first two principal components (\(PC_1\) and \(PC_2\)) using \texttt{gmx anaeig}, defining essential reaction coordinates that describe the dominant collective motions of the system.

\subsection{Free Energy Landscape (FEL)}

The Gibbs free energy surfaces (\(\Delta G\)) were constructed from the joint probability distribution of the projections onto the first two principal components (\(PC_1\) and \(PC_2\)), applying the Boltzmann relation:

\begin{equation}
	\Delta G(PC_1,PC_2) = -k_B T \ln\!\left(\frac{P(PC_1,PC_2)}{P_{\max}}\right) ,
\end{equation}
where \(P(PC_1,PC_2)\) denotes the probability density and \(P_{\max}\) its maximum value, used as the reference energy.
 
\section{Results and Discussion}
\subsection{Validation of the Liquid–Solid Phase Transition}

\begin{figure}[ht!]
 \centering
 \includegraphics{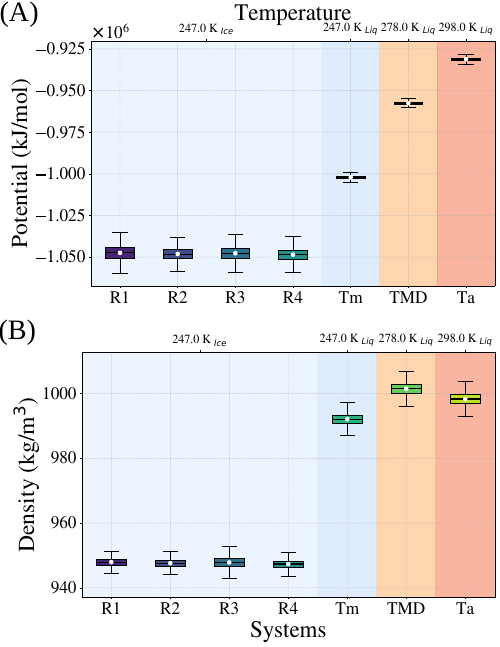}
\caption{
Box plots showing the distributions of system potential energy and solvent density for the simulated systems, extracted from the final 200~ns of the trajectories. Systems are grouped along the $x$ axis according to simulation temperature and solvent phase, explicitly distinguishing the frozen replicas at 247.0~K (\emph{R1--R4}, Ice) from the liquid-water control systems at 247.0~K (\emph{Tm}), 278.0~K (\emph{TMD}), and 298.0~K (\emph{Ta}). 
(A) The potential energy distributions indicate that the liquid systems remain thermodynamically stable across temperatures, whereas the frozen replicas exhibit a pronounced shift toward lower (more negative) energies together with increased variability, consistent with the formation of hexagonal ice (I$_\mathrm{h}$). 
(B) The solvent density distributions reproduce the anomalous behavior of liquid water, with a maximum density at 278.0~K (\emph{TMD}) and a decrease at 247.0~K (\emph{Tm}). In contrast, the frozen replicas display a marked reduction in density to values characteristic of hexagonal ice. Taken together, these observables provide clear and independent evidence that the ice-seeded simulations undergo a liquid--solid phase transition, resulting in frozen solvent environments.
}
 \label{Fig_1}
\end{figure}

Before analyzing the conformational response of Yfh1, we first established that the ice-seeded systems genuinely undergo the liquid–solid transition on which our entire analysis rests. Because the reorganization of the solvent unfolds over hundreds of nanoseconds, distinguishing the stabilized crystalline state from the preceding nucleation and growth regimes is essential: mixing these regimes would blur every downstream observable. We therefore monitored solvent density, potential energy, and both Coulombic and Lennard-Jones interaction energies across the full 1 $\mu$s of each trajectory (Figure \ref{Fig_1S}, Supplementary Information), and restricted all subsequent structural analysis to the final 200 ns, where these observables fluctuate around stationary mean values (Figure \ref{Fig_1}).

Inspection of the full trajectories (Figure~\ref{Fig_1S}) shows that the ice-seeded replicas undergo a clear reorganization in density and potential energy before reaching a quasi-stationary regime, whereas the liquid controls remain stable. To ensure that the reported distributions reflect the stabilized crystalline state{--}and to avoid mixing configurations from the liquid, nucleation, and growth regimes{--}we restricted the analysis to the final 200 ns, corresponding to the post-transition regime where observables fluctuate around stable mean values (Figure~\ref{Fig_1}).

As shown in Figure~\ref{Fig_1}A, the liquid systems exhibit narrowly distributed potential energy values, reflecting thermodynamic stability at 298.0~K (\emph{Ta}), 278.0~K (\emph{TMD}), and 247.0~K (\emph{Tm}). Within these liquid controls, a clear temperature dependence is observed, with progressively lower potential energy values as the temperature decreases. In contrast, the ice-seeded replicas display a distribution shifted toward lower potential energy values, centered around $\sim -1.05 \times 10^{6}$~kJ/mol. Notably, these replicas are simulated at the same temperature as the \emph{Tm} system, yet exhibit lower energies. This indicates that the energetic stabilization cannot be attributed solely to temperature.

Consistently, the solvent density distributions further support the occurrence of freezing in the ice-seeded systems (Figure~\ref{Fig_1}B). While the control simulations reproduce the anomalous density behavior of liquid water{--}with a maximum at 278.0~K (\emph{TMD}){--}replicas \emph{R1--R4} exhibit a pronounced reduction in density to values near $\sim 947$~kg/m$^{3}$, characteristic of hexagonal ice (I$_\mathrm{h}$).

\begin{figure*}[ht!]
 \centering
 \includegraphics[width=1.0\textwidth]{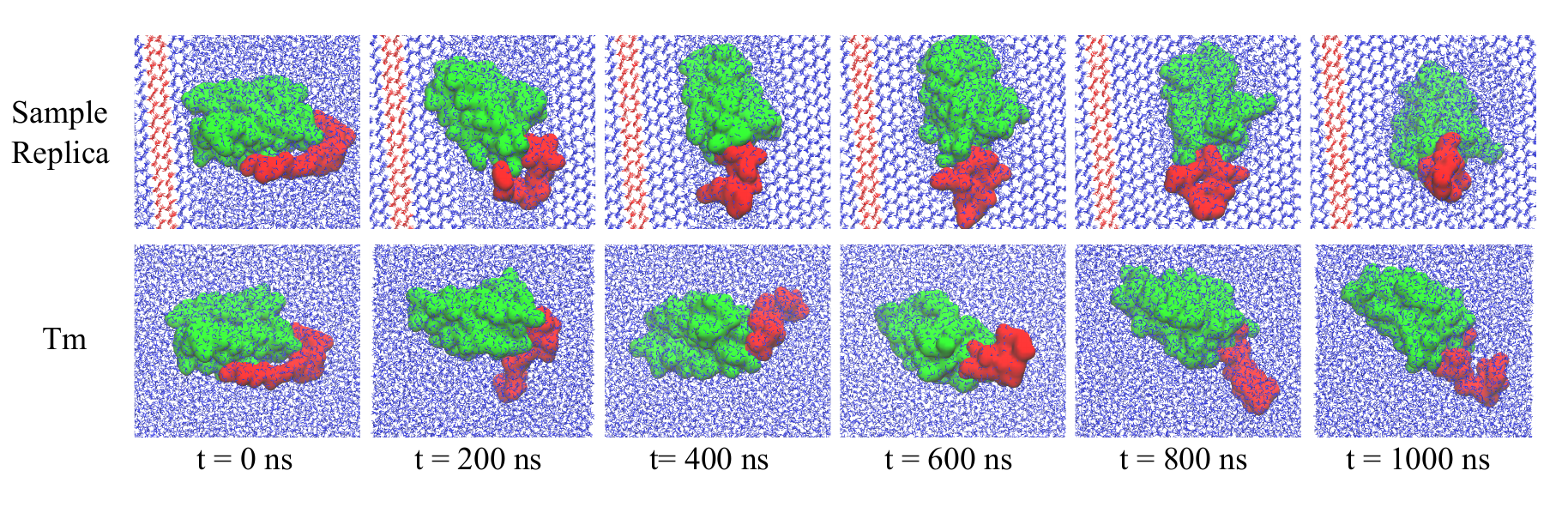}
\caption{
Temporal evolution of frataxin in water with and without a crystallization seed. 
Representative snapshots are shown for one freezing replica and the control simulation at the melting temperature (\emph{Tm}). 
The protein surface is colored in red and green, highlighting the intrinsically disordered N-terminal region (residues 1--19) and the globular domain, respectively. 
In the crystallization replica, water molecules belonging to the ice seed are shown in orange, whereas bulk water molecules are displayed in blue. 
The progressive development of crystalline order is evident in the seeded system, in contrast to the persistent liquid-like arrangement observed in the control simulation.
}
 \label{Fig_2}
\end{figure*}

Direct structural evidence of the liquid--solid transformation is provided in Figure~\ref{Fig_2}, which shows representative snapshots at 200~ns intervals throughout the 1~$\mu$s simulations. These configurations reveal the progressive development and stabilization of an extended ice I$_\mathrm{h}$ lattice in the seeded systems, in clear contrast to the persistent liquid-like arrangement observed in the \emph{Tm} control simulation. Together, the concomitant decrease in potential energy and solvent density, along with the emergence of long-range crystalline order, provides compelling evidence that the seeded simulations undergo a liquid--solid transition toward a stabilized ice I$_\mathrm{h}$ phase.

\begin{figure}[ht!]
 \centering
 \includegraphics[width=0.5\textwidth]{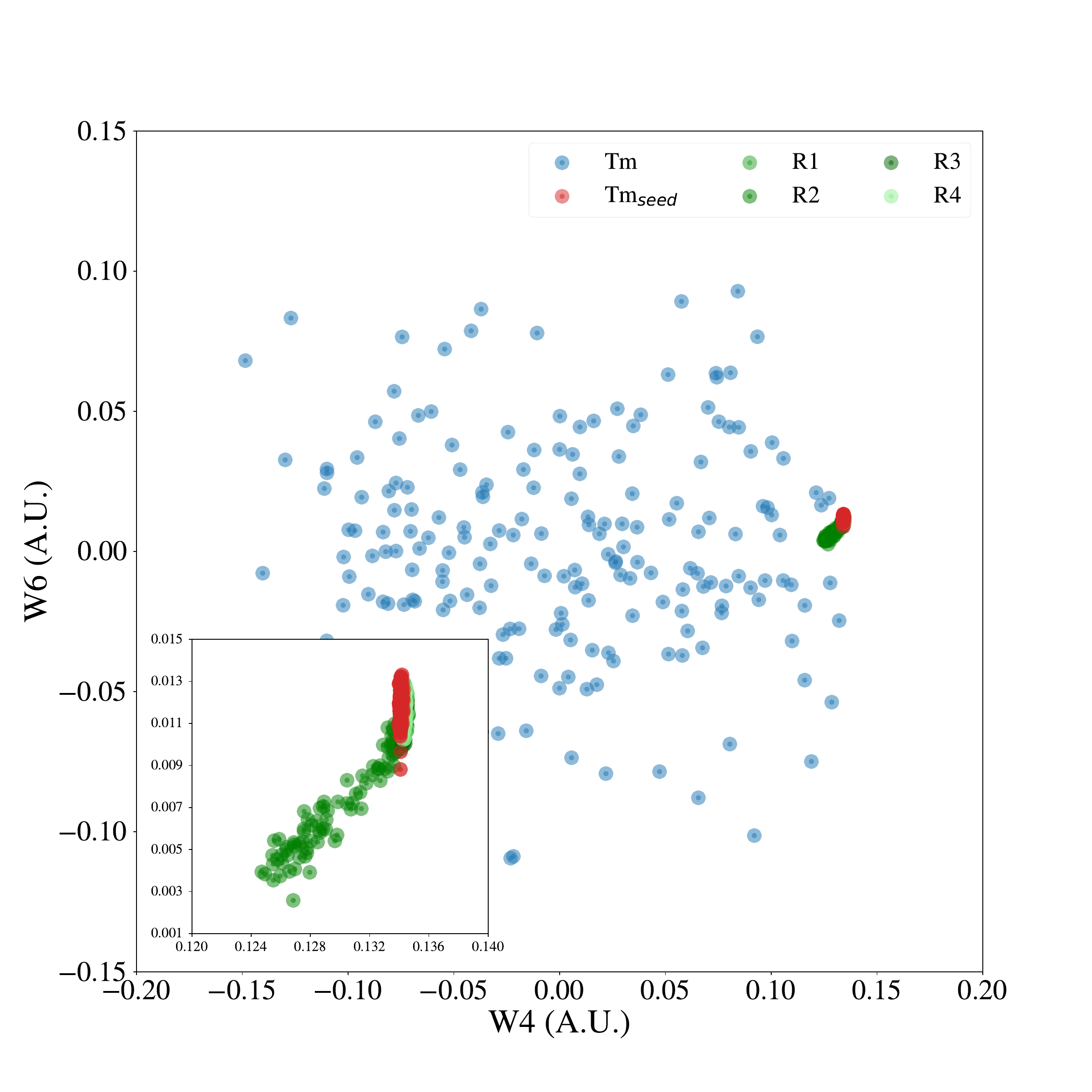}
\caption{Local Bond Order Parameters W6 vs.~W4 in arbitrary units for replicas \emph{R1--R4}, \emph{Tm} and \emph{Tm$_{seed}$} systems, showing a clear liquid behavior for \emph{Tm}, and Ice 1H values for replicas and \emph{Tm$_{seed}$}.}
 \label{Fig_3}
\end{figure}

To finally identify the freezing behavior, we analyzed local Bond Order Parameters W4 and W6 \cite{Steinhardt1983,Wang2005} for water molecules in the last 200 ns of replicas \emph{R1--R4} and compare them with the values obtained for \emph{Tm} and \emph{Tm$_{seed}$} systems. The inspection of Figure \ref{Fig_3} shows that all replicas' water order parameter values evolve, reaching values very close to the ones of \emph{Tm$_{seed}$} system. In addition, a greater variability of the values obtained for these parameters is observed for the \emph{Tm} system, a typical behavior of a system in liquid state, while for the replicas and \emph{Tm$_{seed}$}, the variability is considerably lower.

\begin{figure*}[ht!]
 \centering
 \includegraphics[width=1.0\textwidth]{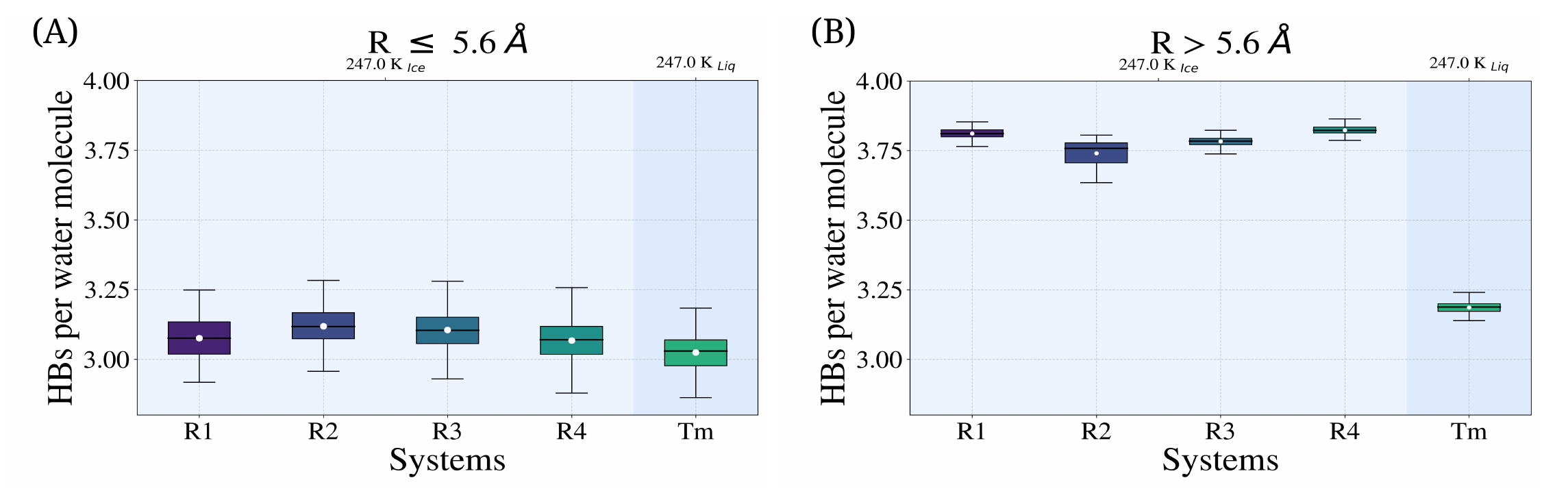}
\caption{Average number of hydrogen bonds per water molecule as a function of the distance from the protein surface. Panel (A) corresponds to water molecules located within 5.6~\AA{} of the protein, representing the first hydration shell, whereas panel (B) includes water molecules located beyond 5.6~\AA{}, corresponding to bulk-like water. Boxplots were calculated from the final 200~ns of each trajectory, corresponding to the quasi-stationary regime. The four freezing replicas (\emph{R1--R4}) correspond to systems in which water crystallized into hexagonal ice (I$_\mathrm{h}$), whereas \emph{Tm} represents the liquid-water control simulated at the same temperature (247.0~K). The shaded regions distinguish ice (\emph{R1--R4}) and liquid (\emph{Tm}) conditions. White circles indicate mean values.}
 \label{Fig_4}
\end{figure*}

To inspect the behavior of water molecules near the protein, we analyzed the average total number of hydrogen bonds per water molecule (Protein-Water and Water-Water), in a region within 5.6~\AA{} (two water molecular radii) of the protein surface. Comparing these values with the ones for water molecules outside of this region (Figure \ref{Fig_4}), it can be seen that for the simulation replicas, the total number of hydrogen bonds per water molecule is very similar to the corresponding value for the liquid system (\emph{Tm}). On the other hand, a marked difference can be seen for water molecules outside this region (R > 5.6~\AA) due to the freezing behavior of the systems.

\subsection{Structural Response of Frataxin under Freezing Conditions}

To compare the global dynamical behavior of frataxin under control and freezing conditions, an exploratory analysis of the sampled conformational space was performed using principal component analysis (PCA) based on C$_\alpha$ atomic positions, complemented by free energy landscapes (FEL) projected onto the first two principal components (PC1–PC2) (See Figure~\ref{Fig_2S}, Supplementary Information).

The PCA reveals clear differences in both the extent and organization of the conformational space sampled under each condition. At room temperature (\emph{Ta}), the protein explores the largest region of the PC1–PC2 subspace, indicating enhanced structural plasticity and the presence of multiple conformational substates. In contrast, the \emph{TMD} condition exhibits a markedly more restricted sampling, despite corresponding to a higher temperature than \emph{Tm}, and closely resembles the behavior observed in the ice replicas (\emph{R1–R4}). Under \emph{Tm} conditions, the protein still samples a relatively broad conformational space; however, the associated free energy landscape appears more compact and homogeneous than in \emph{Ta}, consistent with a smoother distribution of accessible microstates. 
If we compare free energy landscapes for \emph{Ta} and \emph{Tm} conditions (\emph{i.e.}, the broader ones), the first is deeper and exhibits a well-defined funnel-shaped topology, consistent with a more stable conformational state. In contrast, the second landscape is significantly flatter and populated by numerous local minima, indicating a less stable conformation. This shallower and more rugged profile facilitates broader exploration of the conformational free energy landscape, allowing the system to more readily sample multiple states. Therefore, this last condition would make the protein more prone to denaturation.

Finally, the ice I$_\mathrm{h}$ replicas display a pronounced contraction of the conformational landscape, consistent with mechanical confinement imposed by the crystalline ice network.

Overall, these results indicate that (i) protein dynamics is not governed solely by temperature but is also strongly influenced by the structural and dynamical properties of the surrounding water, and (ii) solvent crystallization, together with the anomalous properties of liquid water, reduces the global conformational freedom of frataxin.

To complement the C$_\alpha$-based PCA, which describes the global extent of conformational sampling, we next employed dPCA to examine the internal backbone dynamics of Yfh1. This analysis allows us to assess whether the restricted conformational space observed under freezing conditions is also reflected in the essential dihedral motions of the folded core.

The dPCA analysis was restricted to the globular domain of the protein, excluding residues 1--19 of the intrinsically disordered N-terminal region to prevent stochastic fluctuations of the disordered tail from masking the intrinsic dynamics of the folded core \cite{espinosa2017} (Figure~\ref{Fig_3S}, Supplementary Information). The resulting trajectories were projected onto the subspace defined by the first two principal components (PC1 and PC2), expressed in dimensionless units derived from the trigonometric transformation of the backbone $\phi$ and $\psi$ dihedral angles, enabling a continuous representation of conformational space while avoiding discontinuities inherent to angular variables \cite{mu2005energy}.

\begin{figure*}[!h]
 \centering
 \includegraphics{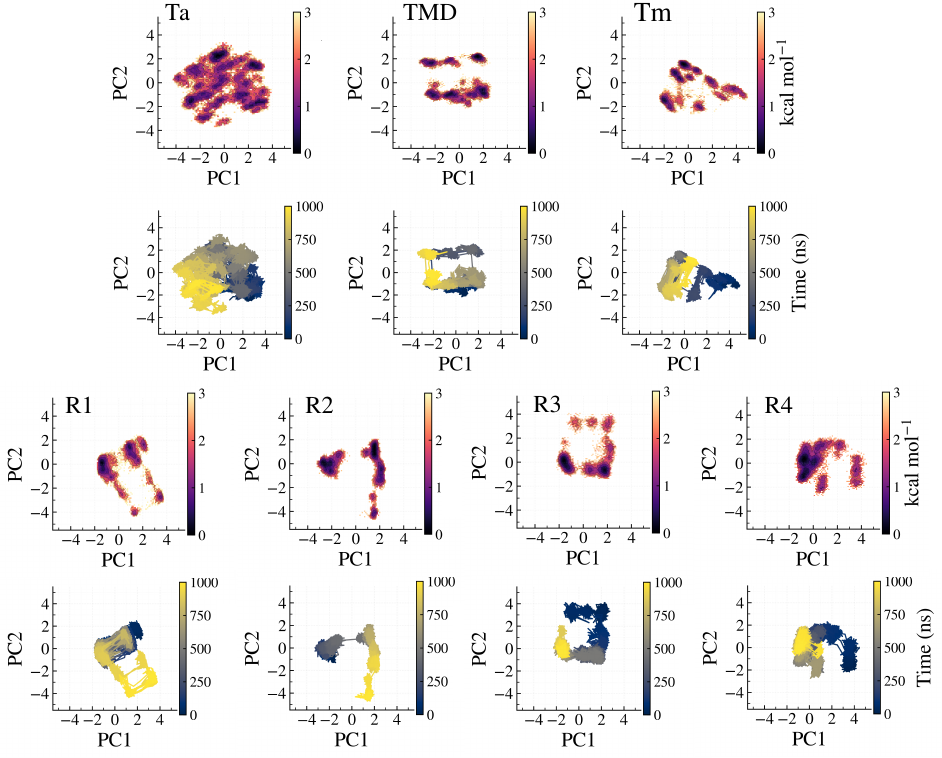}
 \caption{
Dihedral principal component analysis (dPCA) of the conformational space sampled by Yfh1 under control and freezing conditions. The projections correspond to the first two principal components (PC1 and PC2) derived from the backbone $\phi$ and $\psi$ dihedral angles of the globular domain. For each condition, the upper panels show the corresponding free energy landscapes (FEL) projected onto the PC1–PC2 subspace, while the lower panels display the conformational trajectories colored according to simulation time, allowing visualization of the temporal evolution of the system over the 1~$\mu$s simulation. The top three panels (\emph{Ta}, \emph{TMD}, and \emph{Tm}) correspond to control simulations in liquid water at different temperatures: room temperature (\emph{Ta}, 298.0~K), the temperature of maximum density of the model (\emph{TMD}, 278.0~K), and the melting temperature (\emph{Tm}, 249.5~K). The bottom four panels (\emph{R1--R4}) show independent replicas of the freezing process from liquid water to hexagonal ice (I$_\mathrm{h}$). Both the \emph{Tm} simulation and the \emph{R1--R4} replicas were performed in two stages: an initial 500~ns at 249.5~K, followed by an additional 500~ns at 247.0~K.
}
 \label{Fig_5}
\end{figure*}

In Figure \ref{Fig_5}, the conformational trajectories are shown as time-colored projections onto the essential subspace, enabling visualization of the temporal evolution of the system throughout the 1~$\mu$s simulation. These representations are complemented by the corresponding free energy landscapes (FEL). Together, these analyses allow simultaneous assessment of the extent of conformational sampling, the connectivity between states, and the presence of preferred energetic basins under each condition.

In the liquid systems, the protein samples broad and relatively connected regions of conformational space. At room temperature (\emph{Ta}), the landscape is highly extended and continuous, reflecting substantial structural plasticity. In contrast, the \emph{TMD} condition exhibits a more localized and fragmented conformational space, consistent with a more restricted ensemble. Under \emph{Tm} conditions, the system maintains considerable conformational exploration with multiple partially connected minima, indicating that conformational accessibility is not governed solely by temperature.

In the freezing replicas (\emph{R1–R4}), a pronounced reduction in the accessible conformational space is observed, characterized by the presence of discrete and well-separated minima in the free energy landscape (FEL). This fragmentation suggests the emergence of kinetically trapped conformational states, associated with an effective increase in the energy barriers separating substates. Temporal analysis indicates that during the first $\sim$500~ns, significant conformational transitions occur, coinciding with ice formation, followed by progressive confinement within specific regions of the essential subspace.

The consistency between PCA and dPCA results indicates that this restriction is not limited to global motions but also affects the intrinsic backbone dynamics of the globular domain. Overall, these findings demonstrate that water crystallization not only reduces the conformational freedom of the protein but also reorganizes its energy landscape, shifting from a continuous and connected regime to a discretized landscape dominated by confined states.

\subsection{Solvent Exposure, Compactness, and Hydrogen Bonding Analysis}

To further characterize the structural features of the conformational states identified by dPCA, we analyzed solvent exposure, compactness, and protein--solvent interactions through the total solvent-accessible surface area (SASA), radius of gyration ($R_\mathrm{g}$), and hydrogen bonding over the final 200~ns of each trajectory, corresponding to the quasi-stationary regime. This ensures that the reported distributions reflect the stabilized ensembles observed in the dPCA projections.

\begin{figure*}[ht!]
 \centering
 \includegraphics[width=1.0\textwidth]{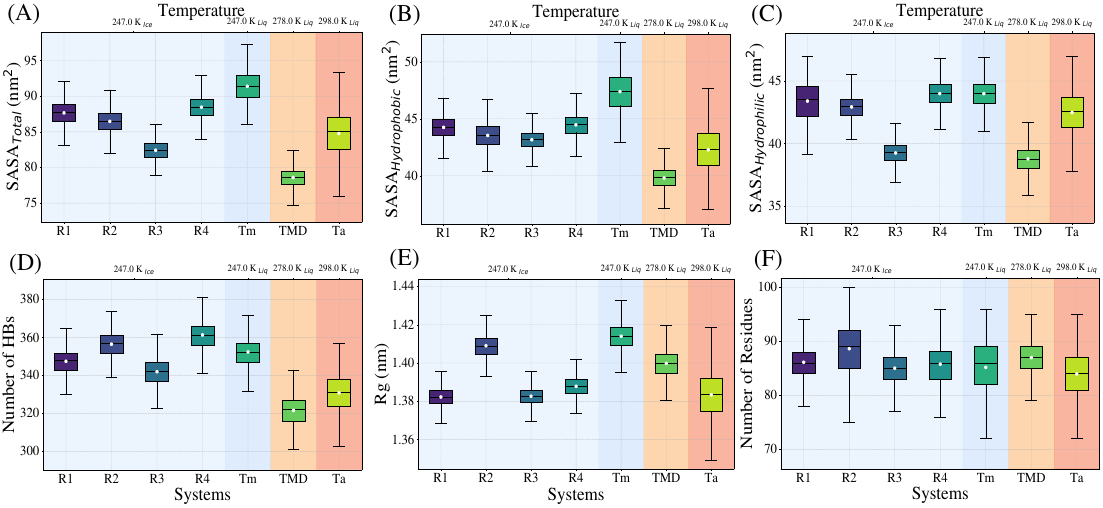}
 \caption{Box plots showing various structural parameters extracted from the final 200 ns of the simulations:
(A) total solvent-accessible surface area ($\text{SASA}_{\text{total}}$),
(B) hydrophobic SASA ($\text{SASA}_{\text{pho}}$),
(C) hydrophilic SASA ($\text{SASA}_{\text{phi}}$),
(D) number of hydrogen bonds (HBs), 
(E) radius of gyration ($\text{R}_{\text{g}}$) and,
(F) number of amino acid residues in the secondary structure.}
 \label{Fig_6}
\end{figure*}

Analysis of the total solvent-accessible surface area ($\text{SASA}_{\text{total}}$) reveals a significant increase under low-temperature liquid conditions (\emph{Tm}) and in the freezing replicas (\emph{R1–R4}) relative to the \emph{Ta} and \emph{TMD} controls. The lowest mean value is observed at \emph{TMD} (78.59~nm\textsuperscript{2}), consistent with the compact conformational ensemble identified in the PCA and dPCA analyses. In contrast, the highest $\text{SASA}_{\text{total}}$ occurs at \emph{Tm} (91.39~nm\textsuperscript{2}), reflecting enhanced structural expansion and broad conformational sampling. The freezing replicas (\emph{R1–R4}) also exhibit increased $\text{SASA}_{\text{total}}$, although to a lesser extent than \emph{Tm}. This intermediate behavior is consistent with the dPCA results, which show that the protein undergoes conformational rearrangements during ice formation but subsequently becomes confined within restricted regions of conformational space. Thus, while partial expansion occurs, the crystalline ice lattice imposes a mechanical constraint that limits further conformational opening (Figure~\ref{Fig_6}A).

Decomposition of SASA into hydrophobic and hydrophilic contributions further supports this interpretation. The freezing replicas display increased nonpolar surface exposure (43.16 – 44.46~nm\textsuperscript{2}) compared to \emph{Ta} and \emph{TMD} (39.80–42.28~nm\textsuperscript{2}), indicating partial exposure of the hydrophobic core consistent with a partially unfolded ensemble. However, these values remain below those observed at \emph{Tm}, reinforcing the notion that structural expansion is modulated by ice-induced confinement (Figure~\ref{Fig_6}B). The hydrophilic component exhibits a modest increase in \emph{Tm} and in most freezing replicas, accompanied by a higher number of protein--water hydrogen bonds (HB). This behavior reflects enhanced accessibility of polar residues and the formation of a more extensive interaction network with the surrounding solvent (Figures~\ref{Fig_6}C and \ref{Fig_6}D). In contrast, the \emph{TMD} system displays the lowest hydrophilic region exposure to solvent, as evidenced by its reduced SASA values (Figures~\ref{Fig_6}A--C), together with the smallest number of protein--water hydrogen bonds (Figure~\ref{Fig_6}D), consistent with a more compact and less hydrated conformational ensemble.

The combined analysis of $\text{SASA}_{\text{total}}$ and the radius of gyration ($\text{R}_{\text{g}}$) further highlights the interplay between expansion and confinement. While \emph{Tm} exhibits the largest $\text{R}_{\text{g}}$ ($\approx$1.414~nm), consistent with maximal structural expansion, the freezing replicas show a more moderate increase in $\text{SASA}_{\text{total}}$ with comparatively smaller changes in $\text{R}_{\text{g}}$. This suggests that, the protein remains globally more compact under ice-confinement conditions than in the supercooled liquid state (Figure~\ref{Fig_6}E).

Finally, the overall secondary-structure content (Figure \ref{Fig_6}F) remains largely preserved across all conditions, indicating that neither low temperature nor solvent crystallization induces major secondary-structure loss. In agreement with the PCA, dPCA, and FEL analyses, these results suggest that the dominant effect of freezing is not global unfolding but rather a restriction and reorganization of the conformational ensemble. Under these conditions, Yfh1 undergoes partial structural expansion followed by confinement, resulting in a reshaped energy landscape characterized by reduced accessibility and stabilized conformational substates.

Importantly, the preservation of the overall secondary-structure content across all conditions is consistent with the known behavior of current biomolecular force fields, which are parameterized to maintain the native fold over the timescales accessible to molecular dynamics simulations. Therefore, the absence of extensive secondary-structure loss does not preclude meaningful conformational changes \cite{adrover2010}. 

\begin{figure}[!h]
 \centering
 \includegraphics{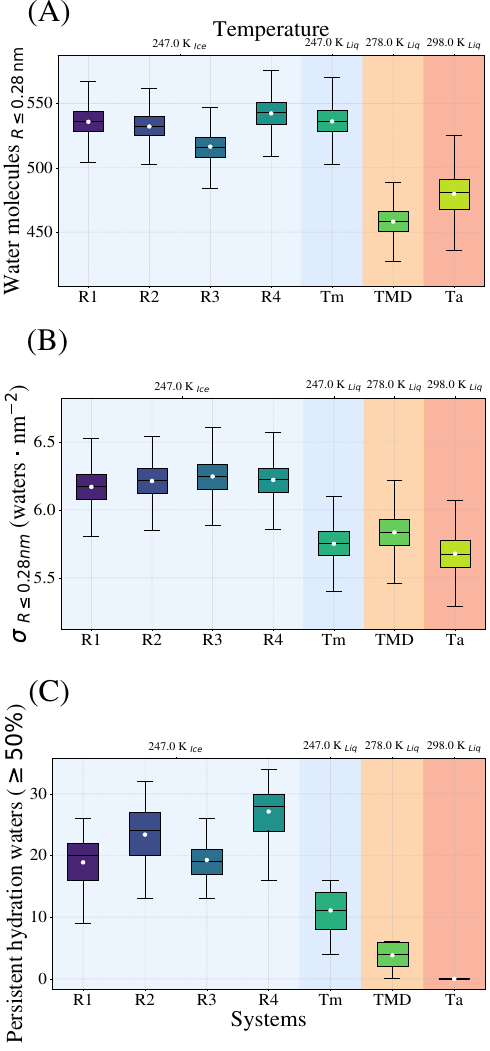}
 \caption{
 Hydration properties of Yfh1 calculated from the final 100~ns of each trajectory, corresponding to the quasi-stationary regime. (A) Total number of water molecules located within 2.8~\AA{} of the protein surface. (B) Surface hydration density, defined as the number of hydration-shell water molecules normalized by the total solvent-accessible surface area (SASA). (C) Persistent hydration waters, corresponding to water molecules that remain within 2.8~\AA{} of the protein surface for at least 50\% of the analyzed trajectory.
 }
 \label{Fig_7}
\end{figure}

To quantify the degree of protein hydration, we defined a surface hydration density ($\sigma$) as the number of water molecules occupying a unit area of the protein surface. The total number of water molecules ($N_{water}$) located within 2.8~\AA{} of the protein surface (Figure~\ref{Fig_7}A) was normalized by the total solvent-accessible surface area (SASA) (Figure~\ref{Fig_6}A). SASA was calculated as the area accessible to a spherical probe with a radius of 1.4~\AA{}, approximately corresponding to the radius of a water molecule (Figure~\ref{Fig_7}B).

\begin{equation}
\sigma = \frac{N_{\mathrm{water}}}{\mathrm{SASA}}
\quad \left[\mathrm{waters} \cdot \mathrm{nm}^{-2}\right]
\end{equation}

To further characterize hydration dynamics, individual water molecules were indexed and their residence times monitored. Water molecules that remained within 2.8~\AA{} of the protein surface for more than 50\% of the analyzed trajectory (\emph{i.e.}, at least half of the last 100 ns of each simulation) were classified as persistent hydration waters (Figure~\ref{Fig_7}C).

Interestingly, the maximum bulk density observed at \emph{TMD} does not translate into a larger number of water molecules surrounding the protein. Instead, the reduced hydration count at \emph{TMD} is primarily due to its lower SASA, which reduces the available interfacial area for solvent accommodation. Consistent with this interpretation, normalization by SASA yields hydration densities comparable to those observed in \emph{Ta} and \emph{Tm}, indicating that the protein is not locally dehydrated at \emph{TMD} but rather adopts a more compact and less solvent-exposed conformational ensemble.

In contrast, the frozen replicas exhibit higher hydration densities despite their reduced conformational freedom. Together with the dPCA and SASA analyses, these results suggest that solvent crystallization promotes the retention of a structured interfacial hydration layer around the protein. Thus, while ice formation restricts conformational sampling and induces confinement of the folded ensemble, it simultaneously favors the persistence of hydration water at the protein surface through the geometrical constraints imposed by the surrounding ice network.

To determine whether this freezing-induced effect was restricted to the immediate protein--water interface or extended into the outer hydration layer, we expanded the cutoff to 5.6~\AA{} (Figure \ref{Fig_5S} in Supplementary Information). At this distance, the effect of freezing became particularly evident in water persistence. While normalization by SASA reduced the differences among the liquid controls, the frozen replicas exhibited a substantially larger population of persistent waters than \emph{Ta}, \emph{TMD}, and \emph{Tm}. This result indicates that the influence of ice formation extends beyond the immediate hydration layer, reducing water exchange and promoting the retention of interfacial waters around Yfh1.

The analyses at 2.8 and 5.6~\AA{} reveal complementary aspects of the hydration response. At \emph{TMD}, the lower absolute hydration count primarily reflects reduced solvent exposure rather than local dehydration, whereas freezing increases the retention of interfacial waters and reduces their exchange dynamics. Thus, protein hydration under these conditions is governed not only by the number of nearby water molecules, but also by the accessible surface area and the persistence of water at the protein--solvent interface.

Overall, these results indicate that the conformational response of Yfh1 under freezing conditions is governed primarily by the structural organization and dynamics of interfacial water, rather than by temperature or bulk water density alone. Although \emph{TMD} corresponds to the maximum bulk density of the TIP4P/2005 water model, it exhibits lower SASA, fewer interfacial water molecules, and fewer protein--water hydrogen bonds, indicating a compact and less solvent-exposed conformational ensemble rather than local dehydration. In contrast, solvent crystallization establishes a distinct interfacial confinement regime. While water beyond the hydration region adopts the ordered structure of ice $I_h$, the interfacial water layer (2.8--5.6~\AA{} from the protein surface) retains liquid-like hydrogen-bonding characteristics, exhibits a higher surface hydration density, and displays markedly longer residence times. This behavior agrees with experimental and computational studies showing that protein stability in frozen systems is largely governed by ice/solution interfaces, freeze-concentrated phases, and quasi-liquid interfacial layers, even in the absence of direct protein--ice contacts \cite{ArsiccioPisano2020,Li2025,Fang2020}. Likewise, simulations and spectroscopic studies of confined proteins have demonstrated that the structure and hydrogen-bond organization of interfacial water are key determinants of conformational stability and dynamics \cite{JinZhuang2025,ReateguiAksan2010,Wang2022}. Together, these findings indicate that the enhanced hydration associated with low temperatures persists under freezing conditions, but its structural consequences are fundamentally altered by solvent crystallization. Rather than promoting unrestricted conformational expansion, the formation of bulk ice reorganizes the interfacial hydration environment, imposing solvent-mediated geometrical constraints that restrict both global and backbone conformational sampling while stabilizing a dense and persistent hydration layer around Yfh1.

\section{Conclusions}

Our study demonstrates that the structural organization of water is a major determinant of protein behavior under freezing conditions. Although water molecules within the interfacial region extending up to $R \leq 5.6$~\AA{} from the protein surface retain a liquid-like hydrogen-bonding pattern, their surface hydration density ($\sigma$) is significantly higher in the freezing replicas than in the liquid controls. This finding indicates that the formation of ice $I_h$ beyond the interfacial hydration region does not displace water from the protein surface, but instead promotes the retention of a denser and more persistent hydration network around the protein.

The increased hydration density is accompanied by a reduction in conformational sampling, as evidenced by the PCA, dPCA, and FEL analyses. In contrast to the \emph{Tm} liquid system, where Yfh1 explores a broader conformational landscape, the freezing replicas exhibit restricted access to conformational substates and a greater tendency toward kinetically confined ensembles. These observations indicate that the crystalline ice network does not directly replace the hydration layer but instead modulates its organization by imposing geometrical constraints on the surrounding solvent environment.

Consequently, the structural organization of bulk water ($R > 5.6$~\AA{}), whether liquid or crystalline, emerges as a key determinant of interfacial hydration and protein dynamics. The transition from the compact ensemble observed at \emph{TMD} to the expanded and highly hydrated states found in \emph{Tm} and the freezing replicas, reflects a shift from a regime dominated by structural stability to one governed by solvent-mediated confinement and kinetic trapping. Under freezing conditions, Yfh1 therefore remains highly hydrated while its conformational landscape becomes increasingly restricted and discretized.

%\balance

%%%REFERENCES%%%
\bibliography{refs} 

@article{simpson1953,
  title={The kinetics of protein denaturation. i. the behavior of the optical rotation of ovalbumin in urea solutions1},
  author={Simpson, Richard B and Kauzmann, W},
  journal={Journal of the American Chemical Society},
  volume={75},
  number={21},
  pages={5139--5152},
  year={1953},
  publisher={ACS Publications}
}

@article{privalov1990,
  title={Cold denaturation of protein},
  author={Privalov, Peter L},
  journal={Critical reviews in biochemistry and molecular biology},
  volume={25},
  number={4},
  pages={281--306},
  year={1990},
  publisher={Taylor \& Francis}
}

@article{tsai2002,
  title={The hydrophobic effect: a new insight from cold denaturation and a two-state water structure},
  author={Tsai, Chung-Jung and Maizel, Jacob V and Nussinov, Ruth},
  journal={Critical reviews in biochemistry and molecular biology},
  volume={37},
  number={2},
  pages={55--69},
  year={2002},
  publisher={Taylor \& Francis}
}

@article{sosnick2025,
  title={Collapse and Protein Folding: Should We Be Surprised that Biothermodynamics Works So Well?},
  author={Sosnick, Tobin R and Baxa, Michael C},
  journal={Annual Review of Biophysics},
  volume={54},
  pages={17–34},
  year={2025},
  publisher={Annual Reviews}
}

@article{adrover2010,
  title={Understanding cold denaturation: the case study of Yfh1},
  author={Adrover, Miquel and Esposito, Veronica and Martorell, Gabriel and Pastore, Annalisa and Temussi, Piero Andrea},
  journal={Journal of the American Chemical Society},
  volume={132},
  number={45},
  pages={16240--16246},
  year={2010},
  publisher={ACS Publications}
}

@article{adrover2012,
  title={The role of hydration in protein stability: comparison of the cold and heat unfolded states of Yfh1},
  author={Adrover, Miquel and Martorell, Gabriel and Martin, Stephen R and Urosev, Dunja and Konarev, Petr V and Svergun, Dmitri I and Daura, Xavier and Temussi, Pierandrea and Pastore, Annalisa},
  journal={Journal of molecular biology},
  volume={417},
  number={5},
  pages={413--424},
  year={2012},
  publisher={Elsevier}
}

@article{sanfelice2016,
  title={Cold denaturation as a tool to measure protein stability},
  author={Sanfelice, Domenico and Temussi, Piero Andrea},
  journal={Biophysical chemistry},
  volume={208},
  pages={4--8},
  year={2016},
  publisher={Elsevier}
}

@article{temussi2025,
  title={Life and death of Yfh1: how cool is cold denaturation},
  author={Temussi, Piero Andrea and Martin, Stephen R and Pastore, Annalisa},
  journal={Quarterly Reviews of Biophysics},
  volume={58},
  pages={e2},
  year={2025},
  publisher={Cambridge University Press}
}

@article{sanfelice2015,
  title={Cold denaturation unveiled: molecular mechanism of the asymmetric unfolding of yeast frataxin},
  author={Sanfelice, Domenico and Morandi, Edoardo and Pastore, Annalisa and Niccolai, Neri and Temussi, Piero Andrea},
  journal={ChemPhysChem},
  volume={16},
  number={17},
  pages={3599--3602},
  year={2015},
  publisher={Wiley Online Library}
}

@article{puglisi2021,
  title={The anatomy of unfolding of Yfh1 is revealed by site-specific fold stability analysis measured by 2D NMR spectroscopy},
  author={Puglisi, Rita and Karunanithy, Gogulan and Hansen, D Flemming and Pastore, Annalisa and Temussi, Piero Andrea},
  journal={Communications chemistry},
  volume={4},
  number={1},
  pages={127},
  year={2021},
  publisher={Nature Publishing Group UK London}
}

@article{puglisi2022,
  title={Heat and cold denaturation of yeast frataxin: The effect of pressure},
  author={Puglisi, Rita and Cioni, Patrizia and Gabellieri, Edi and Presciuttini, Gianluca and Pastore, Annalisa and Temussi, Piero Andrea},
  journal={Biophysical Journal},
  volume={121},
  number={8},
  pages={1502--1511},
  year={2022},
  publisher={Elsevier}
}

@article{alfano2017,
  title={An optimized strategy to measure protein stability highlights differences between cold and hot unfolded states},
  author={Alfano, Caterina and Sanfelice, Domenico and Martin, Stephen R and Pastore, Annalisa and Temussi, Piero Andrea},
  journal={Nature communications},
  volume={8},
  number={1},
  pages={15428},
  year={2017},
  publisher={Nature Publishing Group UK London}
}

@article{hishida2022,
  title={Effect of osmolytes on water mobility correlates with their stabilizing effect on proteins},
  author={Hishida, Mafumi and Anjum, Rubaiya and Anada, Takahisa and Murakami, Daiki and Tanaka, Masaru},
  journal={The Journal of Physical Chemistry B},
  volume={126},
  number={13},
  pages={2466--2475},
  year={2022},
  publisher={ACS Publications}
}

@article{campanile2024,
  title={Is Water the Engine of Protein Folding?},
  author={Campanile, Marco and Graziano, Giuseppe},
  journal={Biophysica},
  volume={4},
  number={4},
  pages={507--516},
  year={2024},
  publisher={MDPI}
}

@article{muller1990,
  title={Search for a realistic view of hydrophobic effects},
  author={Muller, Norbert},
  journal={Accounts of Chemical Research},
  volume={23},
  number={1},
  pages={23--28},
  year={1990},
  publisher={ACS Publications}
}

@article{graziano2005,
  title={On the intactness of hydrogen bonds around nonpolar solutes dissolved in water},
  author={Graziano, Giuseppe and Lee, Byungkook},
  journal={The Journal of Physical Chemistry B},
  volume={109},
  number={16},
  pages={8103--8107},
  year={2005},
  publisher={ACS Publications}
}

@article{graziano2014,
  title={On the mechanism of cold denaturation},
  author={Graziano, Giuseppe},
  journal={Physical Chemistry Chemical Physics},
  volume={16},
  number={39},
  pages={21755--21767},
  year={2014},
  publisher={Royal Society of Chemistry}
}

@article{riccio2011,
  title={Cold unfolding of $\beta$-hairpins: A molecular-level rationalization},
  author={Riccio, Angelo and Graziano, Giuseppe},
  journal={Proteins: Structure, Function, and Bioinformatics},
  volume={79},
  number={6},
  pages={1739--1746},
  year={2011},
  publisher={Wiley Online Library}
}

@article{dias2010,
  title={The hydrophobic effect and its role in cold denaturation},
  author={Dias, Cristiano L and Ala-Nissila, Tapio and Wong-Ekkabut, Jirasak and Vattulainen, Ilpo and Grant, Martin and Karttunen, Mikko},
  journal={Cryobiology},
  volume={60},
  number={1},
  pages={91--99},
  year={2010},
  publisher={Elsevier}
}

@article{dias2012,
  title={Unifying microscopic mechanism for pressure and cold denaturations of proteins},
  author={Dias, Cristiano L},
  journal={Physical review letters},
  volume={109},
  number={4},
  pages={048104},
  year={2012},
  publisher={APS}
}

@article{kuiper2015,
  title={The biological function of an insect antifreeze protein simulated by molecular dynamics},
  author={Kuiper, Michael J and Morton, Craig J and Abraham, Sneha E and Gray-Weale, Angus},
  journal={Elife},
  volume={4},
  pages={e05142},
  year={2015},
  publisher={eLife Sciences Publications Limited}
}

@article{abascal2005,
  title={A general purpose model for the condensed phases of water: TIP4P/2005},
  author={Abascal, Jose LF and Vega, Carlos},
  journal={The Journal of chemical physics},
  volume={123},
  number={23},
  pages = {234505},
  year={2005},
  publisher={AIP Publishing}
}

@article{jorgensen1983,
  title={Comparison of simple potential functions for simulating liquid water},
  author={Jorgensen, William L and Chandrasekhar, Jayaraman and Madura, Jeffry D and Impey, Roger W and Klein, Michael L},
  journal={The Journal of chemical physics},
  volume={79},
  number={2},
  pages={926--935},
  year={1983},
  publisher={American Institute of Physics}
}

@article{conde2017,
  title={High precision determination of the melting points of water TIP4P/2005 and water TIP4P/Ice models by the direct coexistence technique},
  author={Conde, MM and Rovere, M and Gallo, P},
  journal={The Journal of chemical physics},
  volume={147},
  number={24},
  pages={244506},
  year={2017},
  publisher={AIP Publishing}
}

@article{bussi2007,
  title={Canonical sampling through velocity rescaling},
  author={Bussi, Giovanni and Donadio, Davide and Parrinello, Michele},
  journal={The Journal of chemical physics},
  volume={126},
  number={1},
  pages={014101},
  year={2007},
  publisher={AIP}
}

@article{berendsen1984,
  title={Molecular dynamics with coupling to an external bath},
  author={Berendsen, Herman JC and Postma, JPM van and van Gunsteren, Wilfred F and DiNola, ARHJ and Haak, JR},
  journal={The Journal of chemical physics},
  volume={81},
  number={8},
  pages={3684--3690},
  year={1984},
  publisher={AIP}
}

@article{pereyra2011,
  title={Temperature dependence of ice critical nucleus size},
  author={Pereyra, Rodolfo G and Szleifer, Igal and Carignano, Marcelo A},
  journal={The Journal of chemical physics},
  volume={135},
  number={3},
  pages={034508},
  year={2011},
  publisher={AIP}
}

@article{blazquez2022,
  title={Melting points of water models: Current situation},
  author={Blazquez, Samuel and Vega, Carlos},
  journal={The Journal of Chemical Physics},
  volume={156},
  number={21},
  pages={216101},
  year={2022},
  publisher={AIP Publishing}
}

@article{parrinello1981,
  title={Polymorphic transitions in single crystals: A new molecular dynamics method},
  author={Parrinello, Michele and Rahman, Aneesur},
  journal={Journal of Applied physics},
  volume={52},
  number={12},
  pages={7182--7190},
  year={1981},
  publisher={American Institute of Physics}
}

@article{evans1985,
  title={The nose--hoover thermostat},
  author={Evans, Denis J and Holian, Brad Lee},
  journal={The Journal of chemical physics},
  volume={83},
  number={8},
  pages={4069--4074},
  year={1985},
  publisher={American Institute of Physics}
}

@article{he2004,
  title={Yeast frataxin solution structure, iron binding, and ferrochelatase interaction},
  author={He, Yanan and Alam, Steven L and Proteasa, Simona V and Zhang, Yan and Lesuisse, Emmanuel and Dancis, Andrew and Stemmler, Timothy L},
  journal={Biochemistry},
  volume={43},
  number={51},
  pages={16254--16262},
  year={2004},
  publisher={ACS Publications}
}

@article{essmann1995,
  title={A smooth particle mesh Ewald method},
  author={Essmann, Ulrich and Perera, Lalith and Berkowitz, Max L and Darden, Tom and Lee, Hsing and Pedersen, Lee G},
  journal={The Journal of chemical physics},
  volume={103},
  number={19},
  pages={8577--8593},
  year={1995},
  publisher={AIP}
}

@article{harvey2009,
  title={An implementation of the smooth particle mesh Ewald method on GPU hardware},
  author={Harvey, MJ and De Fabritiis, G},
  journal={Journal of chemical theory and computation},
  volume={5},
  number={9},
  pages={2371--2377},
  year={2009},
  publisher={ACS Publications}
}

@article{hess1997,
  title={LINCS: a linear constraint solver for molecular simulations},
  author={Hess, Berk and Bekker, Henk and Berendsen, Herman JC and Fraaije, Johannes GEM},
  journal={Journal of computational chemistry},
  volume={18},
  number={12},
  pages={1463--1472},
  year={1997},
  publisher={Wiley Online Library}
}

@article{best2010,
  title={Protein simulations with an optimized water model: cooperative helix formation and temperature-induced unfolded state collapse},
  author={Best, Robert B and Mittal, Jeetain},
  journal={The journal of physical chemistry B},
  volume={114},
  number={46},
  pages={14916--14923},
  year={2010},
  publisher={ACS Publications}
}

@article{abraham2015,
  title={GROMACS: High performance molecular simulations through multi-level parallelism from laptops to supercomputers},
  author={Abraham, Mark James and Murtola, Teemu and Schulz, Roland and P{\'a}ll, Szil{\'a}rd and Smith, Jeremy C and Hess, Berk and Lindahl, Erik},
  journal={SoftwareX},
  volume={1},
  pages={19--25},
  year={2015},
  publisher={Elsevier}
}

@article{mu2005energy,
  title={Energy landscape of a small peptide revealed by dihedral angle principal component analysis},
  author={Mu, Yuguang and Nguyen, Phuong H and Stock, Gerhard},
  journal={Proteins: Structure, Function, and Bioinformatics},
  volume={58},
  number={1},
  pages={45--52},
  year={2005},
  publisher={Wiley Online Library}
}

@article{altis2007,
  title={Dihedral angle principal component analysis of molecular dynamics simulations},
  author={Altis, Alexandros and Nguyen, Phuong H and Hegger, Rainer and Stock, Gerhard},
  journal={The Journal of chemical physics},
  volume={126},
  number={24},
  pages={244111},
  year={2007},
  publisher={AIP Publishing}
}

@article{espinosa2017,
  title={Essential dynamics of the cold denaturation: pressure and temperature effects in yeast frataxin},
  author={Espinosa, Yanis R and Grigera, J Ra{\'u}l and Caffarena, Ernesto R},
  journal={Proteins: Structure, Function, and Bioinformatics},
  volume={85},
  number={1},
  pages={125--136},
  year={2017},
  publisher={Wiley Online Library}
}

@article{Steinhardt1983,
  title = {Bond-orientational order in liquids and glasses},
  volume = {28},
  ISSN = {0163-1829},
  url = {http://dx.doi.org/10.1103/PhysRevB.28.784},
  DOI = {10.1103/physrevb.28.784},
  number = {2},
  journal = {Physical Review B},
  publisher = {American Physical Society (APS)},
  author = {Steinhardt,  Paul J. and Nelson,  David R. and Ronchetti,  Marco},
  year = {1983},
  month = July,
  pages = {784–805}
}

@article{Wang2005,
  title = {Melting of icosahedral gold nanoclusters from molecular dynamics simulations},
  volume = {122},
  ISSN = {1089-7690},
  url = {http://dx.doi.org/10.1063/1.1917756},
  DOI = {10.1063/1.1917756},
  number = {21},
  journal = {The Journal of Chemical Physics},
  publisher = {AIP Publishing},
  author = {Wang,  Yanting and Teitel,  S. and Dellago,  Christoph},
  pages={214722},
  year = {2005},
  month = June 
}

@article{ArsiccioPisano2020,
  title={The Ice-Water Interface and Protein Stability: A Review},
  author={Arsiccio, Andrea and Pisano, Roberto},
  journal={Journal of Pharmaceutical Sciences},
  volume={109},
  number={7},
  pages={2116--2130},
  year={2020},
  doi={10.1016/j.xphs.2020.03.022},
  publisher={Elsevier}
}

@article{Li2025,
  title={Protein stability and critical stabilizers in frozen solutions},
  author={Li, Jinghan and Lin, Xinhao and Zhen, Zixuan},
  journal={European Journal of Pharmaceutics and Biopharmaceutics},
  volume={214},
  pages={114764},
  year={2025},
  doi={10.1016/j.ejpb.2025.114764},
  publisher={Elsevier}
}

@article{Fang2020,
  title={Stability of Freeze-Dried Protein Formulations: Contributions of Ice Nucleation Temperature and Residence Time in the Freeze-Concentrate},
  author={Fang, Rui and Bogner, Robin H. and Nail, Steven L. and Pikal, Michael J.},
  journal={Journal of Pharmaceutical Sciences},
  volume={109},
  number={6},
  pages={1896--1904},
  year={2020},
  doi={10.1016/j.xphs.2020.02.014},
  publisher={Elsevier}
}

@article{JinZhuang2025,
  title={Protein Confinement Decouples Dynamical Heterogeneity from Structural Preordering in Supercooled Monolayer Water},
  author={Jin, Tan and Zhuang, Wei},
  journal={JACS Au},
  volume={5},
  number={12},
  pages={6254--6264},
  year={2025},
  doi={10.1021/jacsau.5c01219},
  publisher={American Chemical Society}
}

@article{ReateguiAksan2010,
  title={Effects of water on the structure and low/high temperature stability of confined proteins},
  author={Re{\'a}tegui, Eduardo and Aksan, Alptekin},
  journal={Physical Chemistry Chemical Physics},
  volume={12},
  pages={10161--10172},
  year={2010},
  doi={10.1039/C003517C},
  publisher={Royal Society of Chemistry}
}

@article{Wang2022,
  title={Insight into the stability of protein in confined environment through analyzing the structure of water by temperature-dependent near-infrared spectroscopy},
  author={Wang, Shiying and Wang, Mian and Han, Li and Sun, Yan and Cai, Wensheng and Shao, Xueguang},
  journal={Spectrochimica Acta Part A: Molecular and Biomolecular Spectroscopy},
  volume={267},
  pages={120581},
  year={2022},
  doi={10.1016/j.saa.2021.120581},
  publisher={Elsevier}
}
\bibliographystyle{rsc} 

\clearpage
\onecolumn
\section*{Supplementary Material}

\begin{center}

\vspace{0.7cm}

{\Large\bfseries Confinement effects on protein stability in a freezing water environment}

\vspace{0.5cm}

Yanis R. Espinosa$^{a,b}$ H. Ariel Alvarez$^{b,c,d}$, C. Manuel Carlevaro$^{b,e}$

\vspace{0.3cm}

\footnotesize{$^{a}$ Faculty of Basic Sciences, Research Group in Molecular Biology and Genetics (BIOMOGEN), University of Pamplona, Pamplona, Colombia.\\
$^{b}$ Instituto de Física de Líquidos y Sistemas Biológicos (CONICET-UNLP), Calle 59 Nro 789, 1900 La Plata, Argentina.\\
$^{c}$ Instituto de Ciencias de la Salud, Universidad Nacional Arturo Jauretche, Florencio Varela, Argentina.\\
$^{d}$ Departamento de Ciencias Biológicas, Facultad de Ciencias Exactas, Universidad Nacional de La Plata, La Plata, Argentina.\\
$^{e}$ Universidad Tecnológica Nacional, Facultad Regional La Plata, Centro de Investigación en Mecánica Experimental y Computacional, Av. 60 esq. 124 s/n, 1923 Berisso, Argentina.}

\end{center}

\vspace{1cm}

% Reiniciar el contador de figuras
\setcounter{figure}{0}
\renewcommand{\thefigure}{S\arabic{figure}}

% Si también hay tablas
\setcounter{table}{0}
\renewcommand{\thetable}{S\arabic{table}}

% Si también hay ecuaciones
\setcounter{equation}{0}
\renewcommand{\theequation}{S\arabic{equation}}

%%%%%%%%%%%%%%%%%%%%%%%%%%%%%%%%%%%%%%%%%%%%%%%%%%%%%%%%%%

% \usepackage{multirow}

%%%%%%%%%%%%%%%%%%%%%%%%%%%%%%%%%%%%%%%%%%%%%%%%%%%%%%%%%%
\begin{table}[ht!]
\centering
\caption{Step-wise relaxation of atomic position restriction.}
\label{tab:equilibration}
\begin{tabular}{|c|c|c|c|}
\hline
\multicolumn{4}{|c|}{\textbf{Systems: Tm; R1 to R4}} \\
\hline
\textbf{Step} & \textbf{Ensemble} & \textbf{Temperature (K)} & \textbf{Restraint force constants (kJ mol$^{-1}$nm$^{-2}$)} \\
\hline
1 & NVT & 278   & 1000 \\
2 & NVT & 278   & 1000 \\
3 & NpT & 268   & 500  \\
4 & NpT & 258   & 250  \\
5 & NpT & 249.5 & 100  \\
6 & NpT & 249.5 & 50   \\
\hline
\multicolumn{4}{|c|}{\textbf{Systems: Ta and TMD}} \\
\hline

\hline
1 & NVT & 278 & 1000 \\
2 & NVT & 278 & 1000 \\
3 & NpT & 278 & 500  \\
4 & NpT & 278 & 250  \\
5 & NpT & 278 & 100  \\
6 & NpT & 278 & 50   \\
\hline
\end{tabular}
\end{table}

%%%%%%%%%%%%%%%%%%%%%%%%%%%%%%%%%%%%%%%%%%%%%%%%%%%%%%%%%%

\begin{figure}
\centering
\includegraphics[
    width=\textwidth,
    height=0.85\textheight,    
]{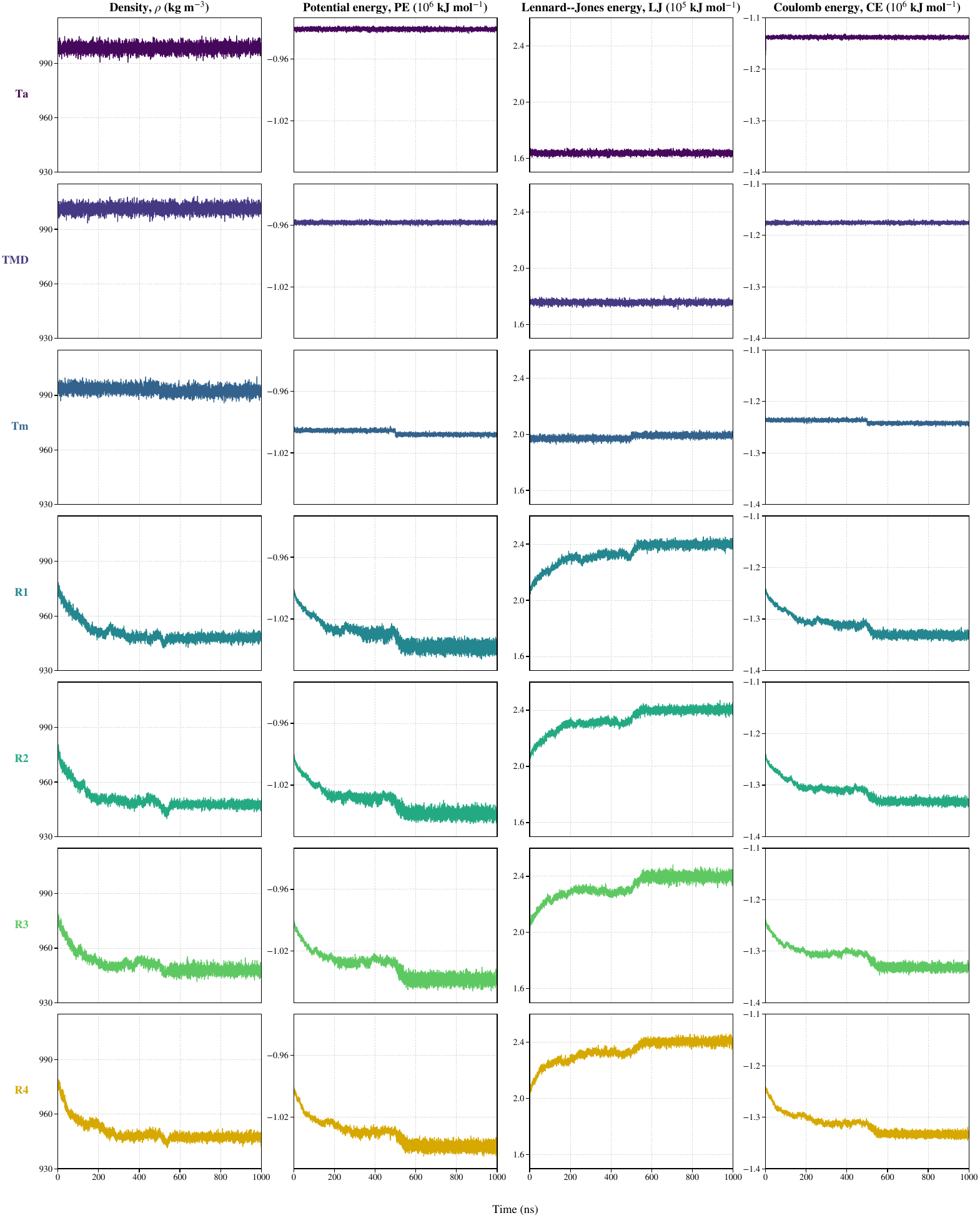}
\caption{Time evolution of density ($\rho$), potential energy (PE), Coulomb energy (CE), and Lennard-Jones energy (LJ) over a 1~$\mu$s simulation. The top three panels (\emph{Ta}, \emph{TMD}, and \emph{Tm}) correspond to control simulations of liquid water at different temperatures: room temperature (\emph{Ta}, 298.0~K), the model's temperature of maximum density (\emph{TMD}, 278.0~K), and the model's melting temperature (\emph{Tm}, 249.5~K). The bottom four panels (\emph{R1--R4}) show independent replicas of the freezing process from liquid water to hexagonal ice (I$_\mathrm{h}$). Both the \emph{Tm} simulation and the \emph{R1--R4} replicas were performed in two stages: an initial 500~ns at 249.5~K followed by 500~ns at 247.0~K.}
\label{Fig_1S}
\end{figure}

%%%%%%%%%%%%%%%%%%%%%%%%%%%%%%%%%%%%%%%%%%%%%%%%%%%%%%%%%%%

%%%%%%%%%%%%%%%%%%%%%%%%%

\begin{figure}
\centering
\includegraphics[width=1.0\textwidth]{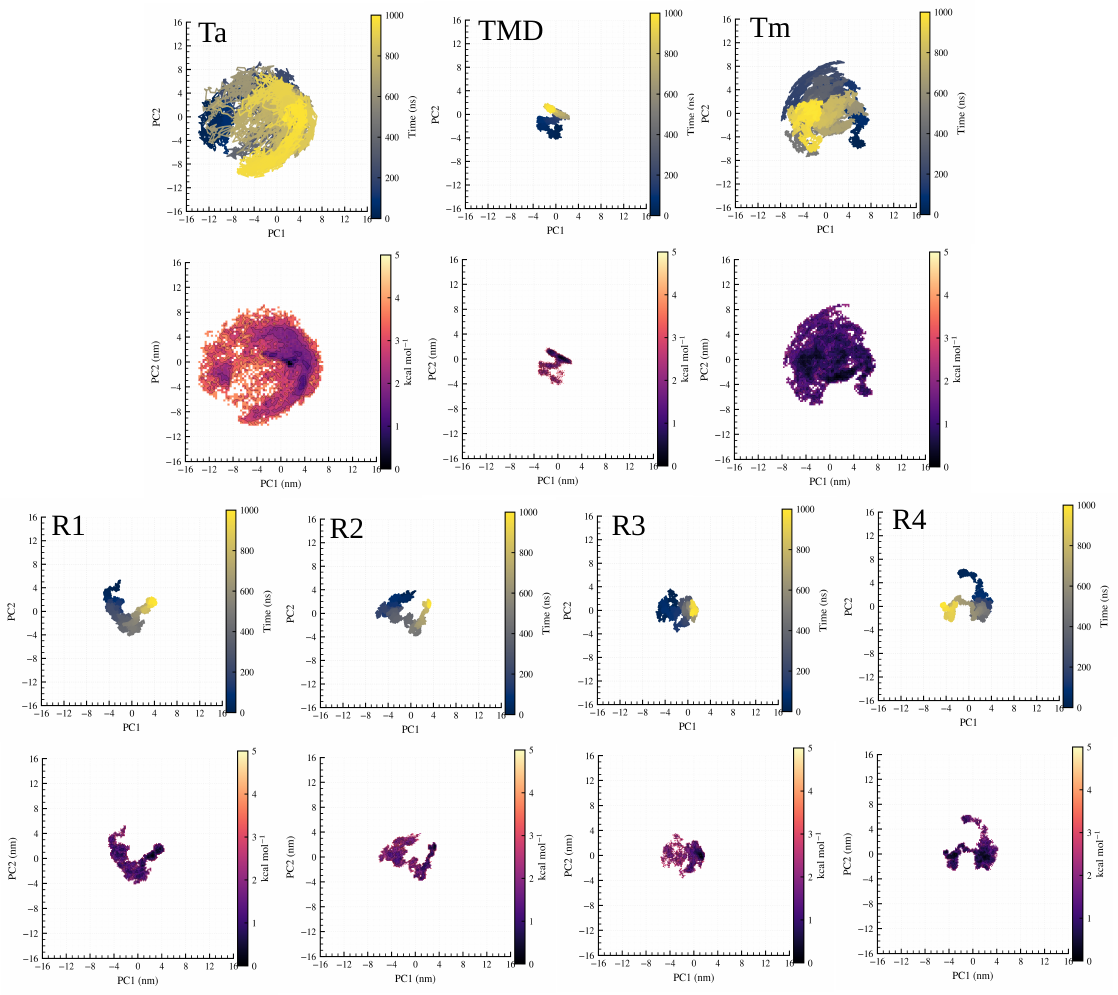}
\caption{Principal component analysis (PCA) of the conformational space sampled by frataxin under control and freezing conditions. The projections correspond to the first two principal components (PC1 and PC2) calculated from the C$_\alpha$ atomic positions. For each condition, the upper panel shows the trajectory projected onto the essential subspace, colored according to simulation time, while the lower panel presents the corresponding free energy landscape (FEL). The top three panels (\emph{Ta}, \emph{TMD}, and \emph{Tm}) correspond to control simulations in liquid water at different temperatures: room temperature (\emph{Ta}, 298.0~K), the temperature of maximum density of the model (\emph{TMD}, 278.0~K), and the melting temperature (\emph{Tm}, 249.5~K). The bottom four panels (\emph{R1--R4}) show independent replicas of the freezing process from liquid water to hexagonal ice (I$_\mathrm{h}$). Both the \emph{Tm} simulation and the \emph{R1--R4} replicas were performed in two stages: an initial 500~ns at 249.5~K, followed by an additional 500~ns at 247.0~K. The comparison reveals clear differences in the extent and organization of the conformational space sampled under each condition, highlighting the impact of temperature and solvent phase on the global dynamics of frataxin.}
\label{Fig_2S}
\end{figure}

%%%%%%%%%%%%%%%%%%%%%%%%%%%%%%%%%%%%%%%%%%%%%%%%%%%%%%%%%%%

\begin{figure}
\centering
\includegraphics[width=1.0\textwidth]{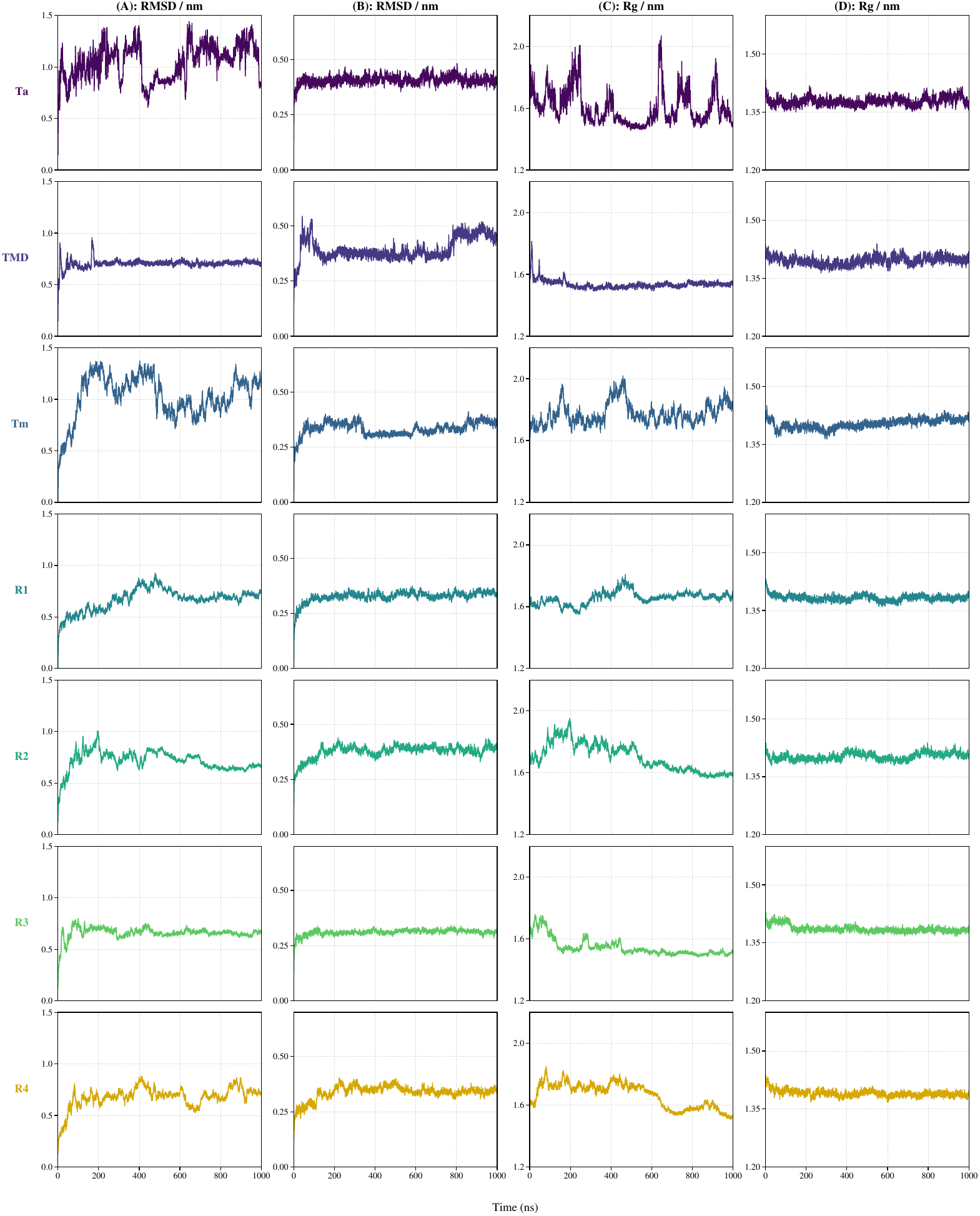}
\caption{Temporal evolution of the root mean square deviation (RMSD) and the radius of gyration ($R_\mathrm{g}$) over a 1~$\mu$s simulation. Panels (A) and (C) show the RMSD and $R_\mathrm{g}$ calculated for all protein atoms, whereas panels (B) and (D) present the corresponding quantities after excluding residues 1--19, allowing a clearer assessment of the fluctuation overestimation arising from the intrinsically disordered N-terminal region. The top three panels (\emph{Ta}, \emph{TMD}, and \emph{Tm}) correspond to control simulations of liquid water at different temperatures: room temperature (\emph{Ta}, 298.0~K), the model's temperature of maximum density (\emph{TMD}, 278.0~K), and the model's melting temperature (\emph{Tm}, 249.5~K). The bottom four panels (\emph{R1--R4}) show independent replicas of the freezing process from liquid water to hexagonal ice (I$_\mathrm{h}$). Both the \emph{Tm} simulation and the \emph{R1--R4} replicas were performed in two stages: an initial 500~ns at 249.5~K followed by 500~ns at 247.0~K.}
\label{Fig_3S}
\end{figure}

%%%%%%%%%%%%%%%%%%%%%%%%%%%%%%%%%%%%%%%%%%%%%%%%%%%%%%%%%%%

\begin{figure}
\centering
\includegraphics[width=1.0\textwidth]{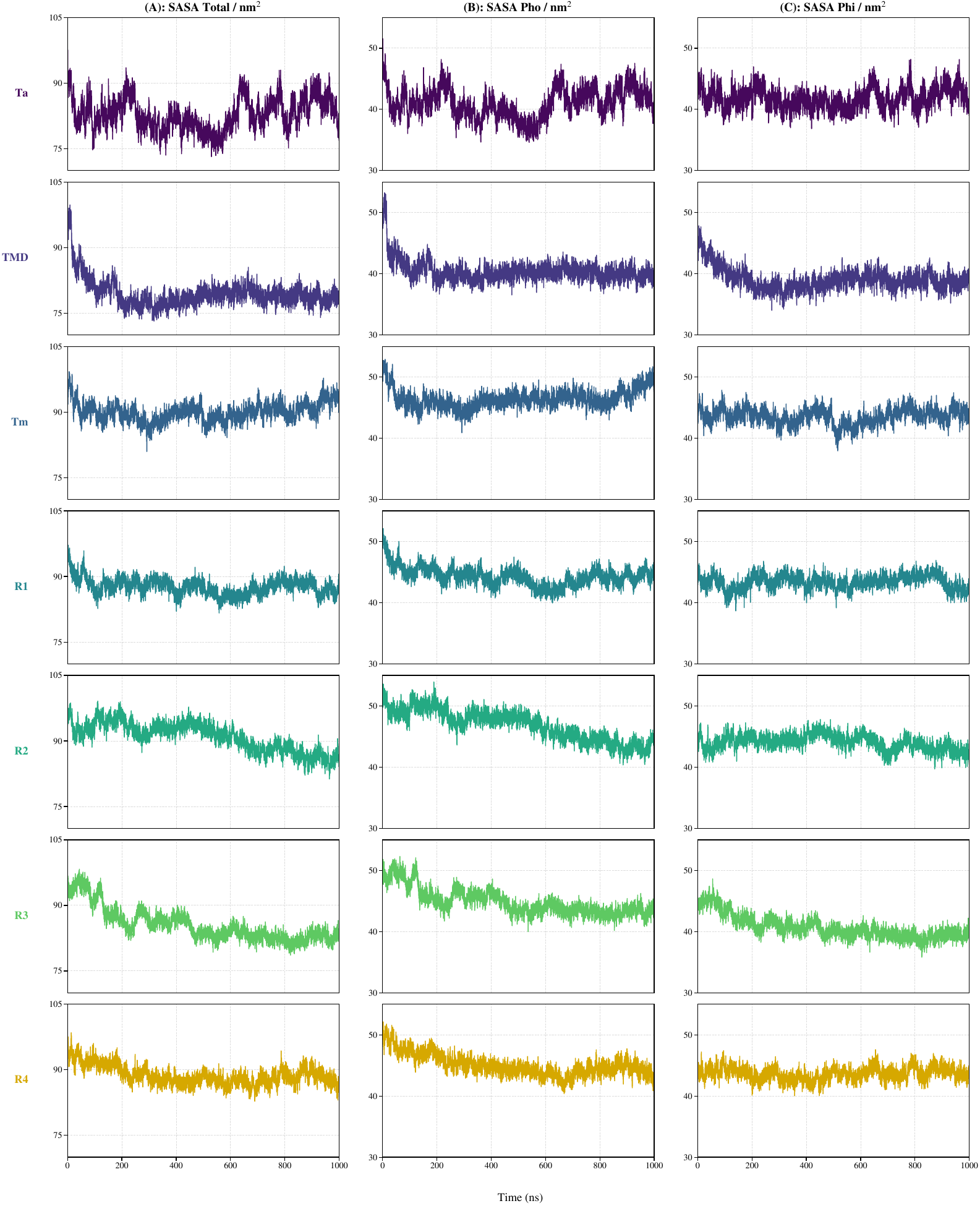}
\caption{ Temporal evolution of the solvent-accessible surface area (SASA) over a 1~$\mu$s simulation. Panels (A), (B), and (C) show the total SASA ($\text{SASA}_{\text{total}}$), hydrophobic SASA ($\text{SASA}_{\text{pho}}$), and hydrophilic SASA ($\text{SASA}_{\text{phi}}$), respectively. The top three panels (\emph{Ta}, \emph{TMD}, and \emph{Tm}) correspond to control simulations of liquid water at different temperatures: room temperature (\emph{Ta}, 298.0~K), the model's temperature of maximum density (\emph{TMD}, 278.0~K), and the model's melting temperature (\emph{Tm}, 249.5~K). The bottom four panels (\emph{R1--R4}) show independent replicas of the freezing process from liquid water to ice I$_\mathrm{h}$. Both the \emph{Tm} simulation and the \emph{R1--R4} replicas were run in two stages: the first 500~ns at 249.5~K, followed by 500~ns at 247.0~K.}
\label{Fig_4S}
\end{figure}

%%%%%%%%%%%%%%%%%%%%%%%%%%%%%%%%%%%%%%%%%%%%%%%%%%%%%%%%%%%

\begin{figure}
\centering
\includegraphics[width=0.6\textwidth]{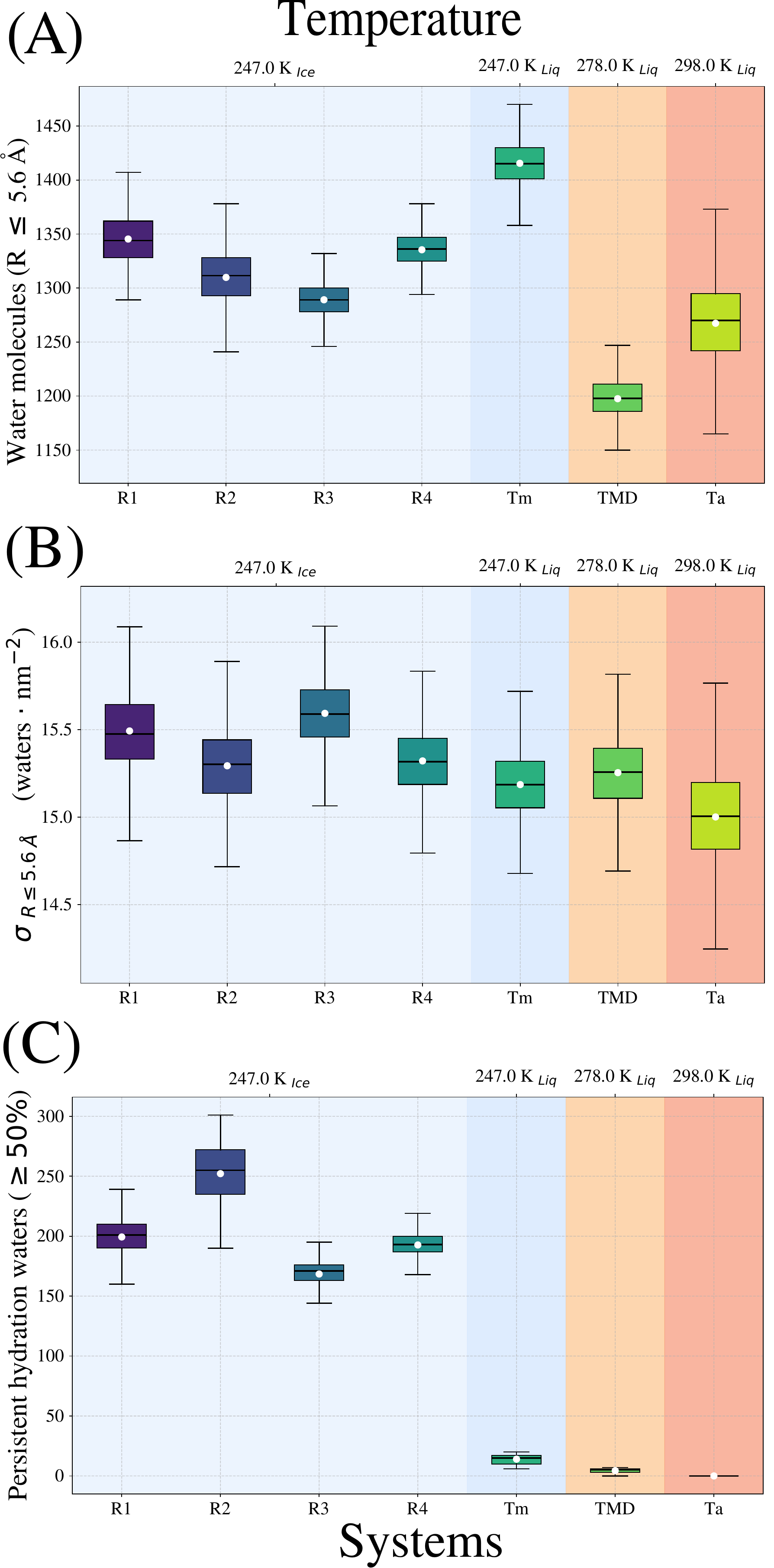}
\caption{  Hydration properties of Yfh1 calculated from the final 100~ns of each trajectory, corresponding to the quasi-stationary regime. (A) Total number of water molecules located within 5.6~\AA{} of the protein surface. (B) Surface hydration density, defined as the number of hydration-shell water molecules normalized by the total solvent-accessible surface area (SASA). (C) Persistent hydration waters, corresponding to water molecules that remain within 5.6~\AA{} of the protein surface for at least 50\% of the analyzed trajectory.}
\label{Fig_5S}
\end{figure}

%%%%%%%%%%%%%%%%%%%%%%%%%%%%%%%%%%%%%%%%%%%%%%%%%%%%%%%%%%



%% References

%%\section*{References}
%%\bibliographystyle{model1-num-names}
%%\bibliography{bib.bib}






% \end{thebibliography}

\end{document}